# On the dipole approximation in the theory of photoionization of the atom

Alexey N. Hopersky, Alexey M. Nadolinsky, Rustam V. Koneev, *and* Julia N. Tolkunova

*Rostov State Transport University, 344038, Rostov-on-Don, Russia*

*E-mail*: qedhop@mail.ru, amnrnd@mail.ru, koneev@gmail.com, juli_tn@mail.ru

**Abstract.** In the nonrelativistic many-electron approximation of the theory of photoionization of the atom in the formalism of secondary quantization and the theory of irreducible tensor operators, analytical structures for the quadrupole transition operator and its matrix element are specified. A new form of the nonrelativistic criterion for the applicability of the dipole approximation is formulated as the ratio of the complete cross sections of one-electron photoionization of an atom via the channels of dipole and quadrupole transitions.

## Introduction

One of the fundamental operators of nonrelativistic quantum mechanics is the operator of the interaction of an electromagnetic field with an atom [1]:

$$\hat{R} = -\frac{e}{m_e c}\sum_{n=1}^{N}\left(\hat{p}_n \cdot \hat{A}_n\right). \qquad (1)$$

The linear (by field) operator (1) describes, in particular, the process of single photoionization of the atom, which is fundamental in the microcosm. A mathematical problem arises of concretizing and taking into account the complete analytic structure of operator (1) in the formalism of secondary quantization and irreducible tensor operators. In (1), the field operator can be represented in the secondary quantization formalism [2–4] as a plane wave expansion,

$$\hat{A}_n = \sum_{\vec{k}}\sum_{p=1,2}\vec{e}_{\vec{k}p}\left(\hat{a}^{+}_{\vec{k}p}e^{-i\eta_n} + \hat{a}^{-}_{\vec{k}p}e^{i\eta_n}\right), \qquad (2)$$

and in the form of a superposition of multipoles by decomposition of plane waves by vector spherical functions [5, 6]. In (1), (2) $e(m_e)$ – charge (mass) of an electron, $c$ – the speed of light in a vacuum, $N$ – number of electrons in an atom, $\hat{A}_n$ – electromagnetic field operator $\eta_n = (\vec{k}\cdot\vec{r}_n)$, $\vec{k}(\vec{e}_{\vec{k}p})$ – the wave vector (polarization vector) of the photon, $p$ – polarization index, $(\vec{k}\cdot\vec{e}_{\vec{k}p}) = 0$, $\hat{p}_n(\vec{r}_n)$ – momentum operator (radius-vector) $n$–electron atom and $\hat{a}^{+}_{\vec{k}p}$ ($\hat{a}^{-}_{\vec{k}p}$) – photon production (annihilation) operator. For our purpose, which is mathematically more compact for the implementation of the methods of the theory of irreducible tensor operators (quantum theory of angular momentum) [7–10], formula (2) with the decomposition of a plane wave into an alternating power series is presented ($\alpha$ – whole):

$$\exp(i\eta_n) = \sum_{\alpha=0}^{\infty}\frac{1}{\alpha!}(i\eta_n)^{\alpha}. \qquad (3)$$

We will limit ourselves to considering only the values of α = 0,1 in (3) (see **Remark 2**). Then (1) takes the form:

$$\hat{R} \approx \hat{R}_D + \hat{R}_Q, \qquad (4)$$

where $\hat{R}_D$ – dipole ($D$; $\alpha = 0$) and $\hat{R}_Q$ –quadrupole ($Q$; $\alpha = 1$) operators of transition. Then, for the *complete* (probability of photon disappearance *without* photoelectron registration) cross section of a one-electron photoionization of an atom, we have:

$$\sigma \sim \sum_{\lambda=D,Q} |\langle i | \hat{R}_\lambda | f(\lambda) \rangle|^2, \qquad (5)$$

where $|i\rangle$ ($|f(\lambda)\rangle$) – is the total wave function of the initial (final) state of the "atom $\oplus$ photon" system. Getting analytical structures $\hat{R}_Q$, $\langle i | \hat{R}_Q | f(Q) \rangle$, $\sigma$ in the formalism of secondary quantization and the theory of irreducible tensor operators and, as a result, the new form of the nonrelativistic criterion for the applicability of the dipole approximation in the theoretical description of the process of a one-electron photoionization of an atom formed the content of this article.

**Results**

Consider the equivalent (1) *integral Loudon representation* ([11], see Eq. (5.35), page 169) for a multielectron operator of the interaction of an electric field with an atom:

$$\hat{R} = e \sum_{n=1}^{N} \int_0^1 \left( \vec{r}_n \cdot \hat{E}(\lambda \vec{r}_n) \right) d\lambda. \qquad (6)$$

By virtue of the connection ($t$ is time) $\hat{E}(\vec{r},t) = (-1/c) \partial \hat{A}(\vec{r},t)/\partial t$ [12] field operator in (6) at $t = 0$ it looks like:

$$\hat{E}(\lambda \vec{r}_n) = \frac{1}{c} \cdot i \sum_{\vec{k}p} \omega_k \vec{e}_{\vec{k}p} \left( \hat{a}_{\vec{k}p}^+ e^{-i\lambda \eta_n} - \hat{a}_{\vec{k}p}^- e^{i\lambda \eta_n} \right), \qquad (7)$$

where is the circular frequency of the photon $\omega_k = kc$ ($k = |\vec{k}|$). Substituting (7) in (6) we get:

$$\hat{R} = i \frac{e}{c} \cdot \sum_{n=1}^{N} \sum_{\vec{k}p} \omega_k \left( \vec{e}_{\vec{k}p} \cdot \vec{r}_n \right) \int_0^1 \left( \hat{a}_{\vec{k}p}^+ e^{-i\lambda \eta_n} - \hat{a}_{\vec{k}p}^- e^{i\lambda \eta_n} \right) d\lambda. \qquad (8)$$

Now let's formulate and prove two Statements.

**Statement 1.** The multielectron operator of the electric quadrupole junction has the form:

$$\hat{R}_Q = \frac{4\pi}{15} \left( \frac{2\pi}{3} \right)^{1/2} \cdot \frac{e}{c^2} \sum_{\vec{k}p} \omega_k^2 \left( \hat{a}_{\vec{k}p}^+ + \hat{a}_{\vec{k}p}^- \right) \cdot \Phi, \qquad (9)$$

$$\Phi = \sum_{p=-1}^{1} \sum_{p'=-1}^{1} Y_{1p}^*(\vec{e}_{\vec{k}p}) Y_{1p'}^*(\vec{\epsilon}_k) \chi_{pp'}, \qquad (10)$$

$$\chi_{pp'} = \sum_{m=-2}^{2} (-1)^m \begin{pmatrix} 1 & 1 & 2 \\ p & p' & -m \end{pmatrix} \sum_{n=1}^{N} r_n^2 Y_{2m}(\vec{e}_n), \qquad (11)$$

where unit vectors are defined $\vec{\epsilon}_k = \vec{k}/k$ и $\vec{e}_n = \vec{r}_n/r_n$, $Y_{ab}$ – spherical function (* – complex mate symbol) and $3j$ – Wigner's symbol.

**Proof.** Taking into account (4) of (8) we have:

$$\hat{R}_Q = \frac{1}{2} \cdot \frac{e}{c^2} \sum_{\vec{k}p} \omega_k^2 \left( \hat{a}_{\vec{k}p}^+ + \hat{a}_{\vec{k}p}^- \right) \cdot \Lambda, \qquad (12)$$

$$\Lambda = \sum_{n=1}^{N} \left( \vec{e}_{\vec{k}p} \cdot \vec{r}_n \right) \left( \vec{\epsilon}_k \cdot \vec{r}_n \right). \qquad (13)$$

Let's take into account the mathematical facts of the quantum theory of angular momentum [7]:
(a) decomposition of the dot product of vectors by spherical functions:

$$(\vec{e} \cdot \vec{r}_n) = \frac{4\pi}{3} \sum_{p=-1}^{1}(-1)^p \cdot r_n \cdot Y_{1,-p}(\vec{e}) Y_{1p}(\vec{e}_n), \qquad (14)$$

(b) Clebsch-Gordan decomposition:

$$Y_{l_1 m_1}(\vec{e}_n) Y_{l_2 m_2}(\vec{e}_n) = \sum_{LM} \left(\frac{[l_1][l_2]}{4\pi[L]}\right)^{1/2} \cdot C_{l_1 0 l_2 0}^{L0} \cdot C_{l_1 m_1 l_2 m_2}^{LM} \cdot Y_{LM}(\vec{e}_n), \qquad (15)$$

where $[x] \equiv 2x+1$ and the coefficients of vector addition (Clebsch-Gordan coefficients) are determined:

$$C_{l_1 m_1 l_2 m_2}^{LM} = (-1)^{l_1 - l_2 + M} \cdot \sqrt{[L]} \cdot \begin{pmatrix} l_1 & l_2 & L \\ m_1 & m_2 & -M \end{pmatrix}. \qquad (16)$$

Then, substituting (14) in (13), taking into account (15), (16) and the selection rules (for the top line $3j$ – Wigner symbol) $l_1 + l_2 + L = 2g$, $g$ – whole, $L \geq |l_1 - l_2|$, $L \leq l_1 + l_2$, we get (9). By virtue of the selection rules, at $l_1 = l_2 = 1$ quantum number $L$ in (15) takes only the values of $L = 0, 2$. However, the meaning $L = 0$ leaves only a spherical function non-zero $Y_{00}(\vec{e}_n) = 1/\sqrt{4\pi}$. As a result, the corresponding matrix element $\langle i | \hat{R}_Q | f(Q) \rangle = 0$ due to the orthogonality of the spherical parts of wave functions $|i\rangle$ – and $|f(Q)\rangle$ – states. <u>Statement 1 has been proven.</u>

Expression (9) – general analytical result for calculating the probability amplitudes of quadrupole transitions of type $nl \to \varepsilon(l+2)$, $n \leq F$ (Fermi level). In this article, in order to avoid general and cumbersome constructions, we will specify the matrix element $\langle i | \hat{R}_Q | f(Q) \rangle$ for a quadrupole transition from $1s^2$ – shells of an atom with $^1S_0$ – ground state term: $1s^2(^1S_0) \to 1s\varepsilon d(^1D_2)$, $n = 1$, $l = 0$, $\varepsilon$ – the energy of a photoelectron of a continuous spectrum. Populated shells for transition state configurations are not specified hereinafter.

**Statement 2.** Matrix element $\hat{R}_Q$ – operator (9) by quadrupole junction states $|i\rangle = |1s^2(^1S_0)\rangle \otimes \hat{a}_\omega^+ |O_q\rangle$ and $|f(Q)\rangle = |1s\varepsilon d(^1D_2)\rangle \otimes |O_q\rangle$ ($|O_q\rangle$ – the wave function of the photon vacuum of quantum electrodynamics [4]) has the form:

$$\langle i | \hat{R}_Q | f(Q) \rangle = \beta(\omega) \sum_{p=-1}^{1} \sum_{p'=-1}^{1} \begin{pmatrix} 1 & 1 & 2 \\ p & p' & M \end{pmatrix} Y_{1p}^*(\vec{e}_\omega) Y_{1p'}^*(\vec{e}_\omega), \qquad (17)$$

$$\beta(\omega) = \frac{8}{15} \cdot \frac{e}{c} \cdot \left(\frac{\hbar}{V}\right)^{1/2} \cdot (\pi\omega)^{3/2} \cdot \langle 1s | r^2 | \varepsilon d \rangle, \qquad (18)$$

$$\langle 1s | r^2 | \varepsilon d \rangle = \int_0^{+\infty} P_{1s}(r) \cdot r^2 \cdot P_{\varepsilon d}(r) dr. \qquad (19)$$

In (17) $M = 0, \pm 1, \pm 2$ (total moment projection $J = 2$ finite $|1s\varepsilon d(^1D_2)\rangle$ – states of the atom). In (18), (19) $\hbar$ – Planck's constant, $V$ – volume of quantization of the electromagnetic field, $\omega$ – circular frequency of the absorbed photon, $\varepsilon = \hbar\omega - I_{1s}$, $I_{1s}$ – ionization threshold energy of the deep $1s^2$ - shell and $P_{1s}(P_{\varepsilon d})$ – radial part of the wave function of $1s(\varepsilon d)$ – electron.

**Proof.** In (11) enter the notation:

$$\left(\frac{4\pi}{5}\right)^{1/2} \cdot \sum_{n=1}^{N} r_n^2 \cdot Y_{2m}(\vec{e}_n) = Q_m^{(2)}. \tag{20}$$

In approaching $LS$ – communication ($L$ – and $S$ – total orbital momentum and spin of the atom state) take into account the result of the implementation of the Wigner–Eckart theorem [7] for the matrix element of operator (20):

$$\langle 1s^2(^1S_0)|Q_m^{(2)}|1s\varepsilon d(^1D_2,M)\rangle = \begin{pmatrix} 0 & 2 & 2 \\ 0 & m & M \end{pmatrix} (1s^2,{}^1S_0 \| Q^{(2)} \| 1s\varepsilon d,{}^1D_2). \tag{21}$$

In (21), consider the values of $3j$ – Wigner symbol ($\delta_{ab}$ – symbol of Kronecker-Weierstrass) [10]:

$$\begin{pmatrix} 0 & 2 & 2 \\ 0 & m & M \end{pmatrix} = \frac{1}{\sqrt{5}} \cdot (-1)^M \cdot \delta_{m,-M}, \tag{22}$$

and the reduced matrix element ([13], see Eq. (29.5), page 260):

$$(1s^2,{}^1S_0 \| Q^{(2)} \| 1s\varepsilon d,{}^1D_2) = \frac{2}{\sqrt{3}} \cdot \langle 1s|r^2|\varepsilon d\rangle. \tag{23}$$

Let us also take into account the matrix elements of the photon production (annihilation) operators in (9) [1]:

$$\langle \omega | \hat{a}_{\vec{k}p}^{+} | O_q \rangle = \left(\frac{2\pi c^2 \cdot \hbar}{\omega V}\right)^{1/2} \cdot \delta_{\omega \omega_k}, \tag{24}$$

$$\langle \omega | \hat{a}_{\vec{k}p}^{-} | O_q \rangle = 0, \tag{25}$$

$$|\omega\rangle = \hat{a}_{\omega}^{+} | O_q \rangle. \tag{26}$$

Then, substituting (21) into the matrix element of the operator $\chi_{pp'}$ from (11) and considering (24) – (26), for the matrix element you are looking for $\hat{R}_Q$ – operator we get (17). <u>Statement 2 has been proven.</u>

Let us conclude the article with the construction quite obvious of a *new form* of the nonrelativistic criterion for the applicability of the dipole approximation when constructing a complete cross section of photoionization of an atom. Traditionally (see, e.g., [10]), this criterion is defined in the form of (long-wave approximation):

$$\eta = r_{1s}/\lambda_\omega \ll 1, \tag{27}$$

where $\lambda_\omega$ – the wavelength of the absorbed photon and (as an example) $r_{1s}$ – medium radius of the $1s^2$ – shell of the atom. The form of criterion (27) contains a double inequality that leaves the energy area applicability of the criterion *uncertain*. Moreover, the magnitude $\eta$ contains a parameter $r_{1s}$ that is not observed in the experiment and does not contain "observable" - probability amplitudes (matrix elements of operators) of multipole transitions. With the requirement of "observability", "*uncertainty*" practically disappears if the criterion for the applicability of the dipole approximation, using (17), is determined by the ratio of the photoionization cross sections along the channels of the dipole and quadrupole junctions:

$$\xi = \sigma_Q / \sigma_D. \tag{28}$$

Taking into account (5), (17), Fermi's "golden rule" [3, 11, 14], the formula for analytical summing up the products of two $3j$ – Wigner symbols [15] and the formula for the photoionization cross section of the $1s^2$ – shells of an atom with filled shells in the ground state by the dipole junction operator [1] ($\alpha$ – fine structure constant),

$$\sigma_D = \frac{4}{3}\pi^2 \cdot \alpha \cdot \hbar\omega \cdot \langle 1s|r|\varepsilon p\rangle^2, \qquad (29)$$

for $\xi$ we get:

$$\xi = 1{,}7 \cdot 10^{-6} \cdot \left(\hbar\omega \cdot \frac{\langle 1s|r^2|\varepsilon d\rangle}{\langle 1s|r|\varepsilon p\rangle}\right)^2. \qquad (30)$$

In (30), the photon energy and radial integrals are taken as numbers in the atomic system of units. As an example, for a neon-like ion $Fe^{16+}$ at $\hbar\omega = 367.5$ a.e. ($\lambda_\omega = 1.240$ Å), $I_{1s} = 282.9$ a.e. and $r_{1s} = 0.031$ Å get $\xi = 1.4 \cdot 10^{-6}$, whereas $\eta \simeq 2.5 \cdot 10^{-2}$. We see that the result for $\xi$ much more clearly shows the leading role of the dipole transition in photoionization of the $1s^2$ – shells of ion $Fe^{16+}$. *Moreover*, for ion $Fe^{16+}$ in *short*-wave radiation regime, e.g., at $\hbar\omega = 7350$ a.e. ($\lambda_\omega = 0.062$ Å) have $= 10^{-6}$, whereas $\eta \simeq 0.5$ does not satisfy double inequality (27). Thus, the new form of the criterion for the applicability of the dipole approximation (30) when constructing a cross section (5) is also applicable *outside* the long-wave radiation regime. *Of course*, in the case of ionization of valence and subvalent shells of an atom (atomic ions) in the experimental study of, for example, the differential characteristics of photoelectron (his *angular distribution*), we should expect (by virtue of *interference* of probability amplitudes along the channels of dipole and quadrupole transitions for the wave function of a photoelectron, which does not have a definite orbital momentum and its projection, and represented by a functional series of partial waves of multipole $l \in [0; \infty)$ with fixed wave function spin [1, 16, 17]) quite measurable *nondipole* effects (see [18-21] and references therein).

**Remarks**

1. The results obtained, as far as we know, for the first time concretize *nonrelativistic* analytic structures $\hat{R}_Q$, $\langle i | \hat{R}_Q | f(Q)\rangle$ and $\xi$ in the formalism of secondary quantization and the theory of irreducible tensor operators, "*encoded*" in the results of the theory of multipoles [5, 10] when replacing $j_l$ - spherical Bessel functions in $l$-multipole junction operators their decomposition into an endless alternating power series [22].

2. With an increase in the energy of the absorbed photon, we should expect an increase in the role of *relativistic* and *nondipole* effects in determining the values and shape of the total cross-section of the photoionization, first of all, of the intermediate and deep shells of the atom (atomic ions). Analytical accounting of these effects through the functional series (3) with the subsequent implementation of the methods of the theory of irreducible tensor operators is an extremely difficult (if at all solvable) task. The problem of taking into account *relativistic* and *nondipole* effects was formally mathematically solved within the framework of classical electrodynamics (multipole theory [5, 10]), followed by a quantum-mechanical interpretation of the results obtained through the *Bohr correspondence principle* [23, 24]. Thus, it is necessary to calculate the well-known *relativistic* result of multipole theory for the *complete* photoionization cross-section as the (formally mathematically infinite: orbital momentum of electron of final ionization state $l \in [1; \infty)$) sum of the partial cross-sections of multipole transitions ([25], see Eq. (5.1.21), page 297; [26], see page 3; [27], see Eq. (7), page 3055). At the same time, as far as we know, the question of the *analytic* summation of this infinite series and, thus, of the full contribution of *nondipole* transitions, *remains open*. As a result, the question of the contribution of *nondipole* effects in the transition to the *nonrelativistic* limit for this sum *remains open*.

*In this article, we presented the English and extended version of the work* [28].